\documentstyle[12pt]{article}
\advance\textheight by \topskip
\begin{document}
\begin{flushright}
SU-ITP-95-19\\
hep-th/9509102\\
\today\\
\end{flushright}
\vspace{1cm}
\begin{center}
\baselineskip=16pt

{\Large\bf   O(6,22) BPS CONFIGURATIONS\\
\vskip 0.6 cm
 OF THE HETEROTIC STRING \, }  \\

\vskip 2cm

{\bf Klaus Behrndt \footnote{ Permanent address: Institut f\"ur Physik,
Humboldt-Universit\"at, 10115 Berlin, Germany; \\ \hspace*{.25in}E-mail:
behrndt@qft2.physik.hu-berlin.de} and Renata
Kallosh}\footnote { E-mail:
kallosh@physics.stanford.edu}\\
 \vskip 0.2cm
Physics Department, Stanford University, Stanford   CA 94305-4060, USA\\
\vskip .6cm

\vskip 1 cm

\end{center}
\vskip 1 cm
\centerline{\bf ABSTRACT}
\begin{quotation}

We present a static  multi-center magnetic  solution of
toroidally compactified heterotic string theory, which is T-duality covariant.
The space-time geometry
depends on the mass   $M$ and on the O(6,22)-norm  $N$ of the magnetic charges.
For  different range of parameters $(M,N)$-solution includes 1) two
independent positive parameters extremal magnetic black holes  with
non-singular  geometry in stringy frame ($a=1$ black holes included),   2)
$a=\sqrt 3$ extremal black holes,  3) singular massive and massless magnetic
white holes (repulsons). The electric  multi-center solution is
also given in an O(6,22)-symmetric form.
\end{quotation}
\newpage

\baselineskip=15pt
We have found an O(6,22)-covariant (i.e. T-duality covariant) BPS
multi-monopole solution, which solves
the field  equations of the   heterotic string  compactified on a
6-dimensional torus. This solution has  one-half of ${\cal N}=4$ supersymmetry
unbroken. The effective action describes the  ${\cal N}=4$ supergravity
multiplet
interacting with 22~ ${\cal N}=4$ abelian
vector supermultiplets\footnote{ One could as well consider an arbitrary
number of vector multiplets $n$. The symmetry of the  solution would be
O(6,n).}.

We start with  the T-duality invariant bosonic action  in the form of
Maharana-Schwarz \cite{MS}
and  Sen \cite{Sen},
\begin{eqnarray}\label{action}
S &=& {1\over16\pi} \int d^4 x \sqrt{-\det G} \, e^{-2 \phi} \, \Big[ R_G + 4
G^{\mu\nu}
\partial_\mu \phi \partial _\nu\phi +{1\over 8} G^{\mu\nu} Tr(\partial _\mu
{\cal M} L\partial_\nu  {\cal M} L)
\nonumber \\
&& -{1\over 12}  (H_{\mu\nu\rho})^2
 -  G^{\mu\mu'} G^{\nu\nu'} F^a_{\mu\nu} \, (L {\cal M} L)_{ab}
\, F^b_{\mu'\nu'} \Big] \, .
\end{eqnarray}
The  pure magnetic solution is given by

\begin{equation} \label{monopole}
\begin{array}{ccc}
ds^2_{\rm str}= -  dt^2 + e^{-4 U} d\vec{x}^2\ , &&  e^{-4U} = 2 \,\chi^T L
\chi
                                              = e^{4\phi} \ , \\
\nonumber\\
\nonumber \\
{\cal M}= {\bf 1}_{28} + 2 e^{4U} \left( \begin{array} {cc} A^2\, n n^T &  AB\,
np^T
\\
   AB\, p n^T & A^2 \, pp^T \end{array} \right) , & &  \vec{H}_m =
   \partial_m \vec{\chi}\ .
\end{array}
\end{equation}
Here the magnetic potential of the theory  $\vec \chi$ is given by the
28-dimensional  harmonic  $O(6,22)$-vector
\begin{equation}
\vec{\chi}(x) = \left( \begin{array}{c} \vec{\chi}_{\rm vec} (x)\\
\vec{\chi}_{\rm gr}(x)
             \end{array} \right)
	     = \left(\begin{array}{c} A(x) \, \vec{n}  \\ B(x) \, \vec{p}
\end{array}
\right)
	    \ , \qquad
 \partial_i  \partial_i  \vec{\chi}(x) =0 \ ,
\label{magn}\end{equation}
 $A(x)$ and $B(x)$ are harmonic functions, and $\vec n$ and $\vec p$
are arbitrary 22- and 6-dimensional unit vectors respectively. There is one
vector field in each vector supermultiplet and 6 vector fields in supergravity
multiplet.
The
 vector fields of the vector multiplets are described by
$\vec{\chi}_{\rm vec}$,  the 22-dimensional harmonic magnetic potential.  The
6-dimensional harmonic magnetic potential of vector fields from supergravity
multiplet is given by  $\vec{\chi}_{\rm gr}$.

The  28$\times$28 symmetric matrix $L$   with 22 eigenvalues $-1$ and 6
eigenvalues $+1$ defines  the metric in the O$(6, 22)$ space.
The magnetic fields ${H}_m^{(a)} = {1\over 2} \epsilon_{mij} {F}_{ij}^{(a)},~
a= 1, \dots
,28 $, also form an
O$(6, 22)$ vector.
The matrix ${\cal M}$ describes the scalar fields, and there is no axion field.
The canonical metric is
\begin{equation}
ds^2_{\rm can}=  -e^{2 U}dt^2 + e^{-2 U} d\vec{x}^2 \ .
\end{equation}
We can choose the harmonic functions $A$ and $B$ to get a
spherically symmetric solution with asymptotically flat metric and vanishing at
infinity
dilaton and scalar fields ${\cal M}$ as
\begin{equation}
A(x) = \frac{ P_{\rm vec}}{r} \ , \qquad B(x) = {1\over \sqrt 2} + \frac{P_{\rm
gr}}{r} \ .
\end{equation}
Here the  total magnetic charge
of the vector fields of the   22 vector supermultiplets is
\begin{equation}
P_{\rm vec} = \sqrt {(\vec P_{\rm vec})^2 }\ ,\  \qquad \vec P_{\rm vec}\equiv
P_{\rm vec} \,\vec n \ .
\end{equation}
 The total magnetic charge of the vector fields of the supergravity multiplet
is
\begin{equation}
P_{\rm gr} =  \sqrt { (\vec P_{\rm gr})^2 }\ , \qquad \vec P_{\rm gr} \equiv
P_{\rm gr}  \vec p \ .
\end{equation}
The supersymmetric properties of the solution are reflected in the fact that
the mass
is non-negative and is related to the  magnetic charge of the gravitational
multiplet,
\begin{equation} \label{bogom}
 M= |Z|= {1\over \sqrt 2} P_{\rm gr} \geq 0 \ .
\end{equation}
Thus all our monopoles are the BPS states.
The solution is defined by 28 independent parameters
($P_{\rm gr}$, $P_{\rm vec}$, $\vec{n}, \;   \vec{p} $;  $\;$ $\vec{n}^2 =1, \;
\vec{p}^2=1$). The metric and the dilaton depend only on 2 independent
parameters: $M$  and  $P_{\rm vec}$ with $P_{\rm gr}=\sqrt 2 M$.

Since the solution is given in terms of the harmonic functions the
multi-monopole solution
is obtained for the case that each harmonic function has a multi-center form.
In the simplest case of asymptotically flat geometry and vanishing fields we
have
\begin{equation}
A(x) = \sum_s \frac{ P^s_{\rm vec}} {|x-x_s|} \  , \qquad
B(x) = {1\over \sqrt 2} + \sum_s\frac{P_{\rm gr}^s}{|x-x_s|}\ , \qquad  M^s=
{1\over \sqrt 2} P_{\rm gr}^s\ .
\end{equation}

The  observation  crucial for the obtaining of  this extremely simple
multi-monopole solution is the following. The ansatz (\ref{magn}) allows us to
solve the Bianchi identities for purely magnetic solutions keeping
T-duality symmetry manifest. Many magnetic black holes which form particular
cases of this solution were known before, therefore the rest is straighforward.
Indeed,
the metric and the dilaton are scalars with
respect to O(6, 22) and therefore we could simply find them by comparison with
known supersymmetric solutions. In particular, one can  choose different
special bases in the 6 and
22 dimensional space  and check that our solution describes  the S-dual of the
compactified ten-dimensional supersymmetric gravitational waves, generalized
fundamental strings and chiral models \cite{BKO1}. The compactified form of
these solutions is given in \cite{Klaus1}. This set of supersymmetric
solutions has the important property that the values of the  charges of
the supergravity multiplet and of the vector multiplets are independent. In
particular the configuration with the mass  equal to the central charge of the
supergravity multiplet and  smaller than the   charge of the vector
multiplets is supersymmetric. This fact has allowed  to find the massless
electrically charged white  holes (repulsons) \cite{Klaus2}, \cite{K}.

 The ${\cal M}$ matrix could be fixed as well by comparison
of the spherically symmetric case with the
extremal limit of the
solution of Sen \cite{Sen}.
Although he has not given an explicit expression for the magnetic fields but
the asymptotic values of the fields,
the matrix
${\cal M}$ is invariant under S-duality and can be used for comparison with our
solution for
 $M^2\geq {1\over 2}
P_{\rm vec}^2$. The
configurations with
$M^2< {1\over 2}
 P_{\rm vec}^2$, as already explained above, also belong to supersymmetric
magnetic configurations of the heterotic string. This, however, does not follow
directly from the O$(7,23)$  rotation of Kerr solution \cite{Sen} but from the
dimensionally reduced supersymmetric generalized gravitational waves
\cite{BKO1}. The electric partners of these magnetic solutions were identified
in \cite{Klaus2}, \cite{K} with  $N_L=0$ states of the heterotic string.
One can as well use the information available  about the supersymmetric
spherically symmetric solutions of the action (\ref{action}) presented in
\cite{CveticYoum}, \cite{Peet} to verify the one-center case of our solutions.

Let us now discuss the  spherically symmetric magnetic solution in  more
detail. The
conformal factor of the spatial metric in the stringy frame equals
\begin{equation}
  e^{-4U}
 =2\, \chi^T L \chi = 1+ \frac{4M}{r} + \frac{2 (P_{\rm gr}^2 - P_{\rm
vec}^2)}{r^2}\ , \qquad
\sqrt 2 M =P_{\rm gr}\ .
                                             \end{equation}

The causal structure  of the space-time depends dramatically on the relation
between the parameters of the theory, in particular on the relation between the
graviphoton charge $P_{\rm gr}$ and vector multiplets charge $P_{\rm vec}$.
We will use as the second parameter, besides the mass $M$, the O$(6,22)$ norm
of a magnetic charge \begin{equation}
N\equiv {1\over 2} P^a L_{ab} P^b = {1\over 2}(P_{\rm gr}^2 - P_{\rm vec}^2)\ .
\end{equation}
The stringy metric in this notation
 and the dilaton are
\begin{equation}
ds^2_{\rm str} = - dt^2 + { 4N +4Mr + r^2\over r^2}
  \,d\vec{x}^2
\ , \qquad e^{4\phi} = { 4N +4Mr + r^2\over r^2}\ .
\end{equation}
The scalar curvature is
\begin{equation}
R_{\rm str} =   {2(4N)^4 - 32 M N r + 4(6 M^2  -  4N)r^2
\over  (4N  +4Mr + r^2)^3}\ .
\end{equation}
We have calculated also the  square of the Ricci tensor and the square of the
Riemann tensor and have found that there are no new singularities besides those
in
$R_{\rm str} $.
It became clear recently \cite{Klaus2}, \cite{K}  that the negative norm case,
specifically
$N=-1$ which corresponds to $N_L=0$ states of the heterotic string, is
available as a supersymmetric configuration, solving the field equations of the
effective action of the heterotic string. Only the configurations in this class
admit the massless limit while the
configuration remains non-trivial. To study the singularities of our solutions
we have to consider separately
positive, vanishing and negative norm $N$ and positive and vanishing mass $M$.

\begin{itemize}
\item  $M>0, \, N >0$   {\it extremal supersymmetric magnetic black holes,
non-singular in
stringy frame}.

The magnetic charge of the graviphoton exceeds the magnetic charge of the
vector multiplets, the norm is positive. The canonical metric, when considered
in a limit from non-extremal black hole, has a singular horizon.
The stringy space-time, however,  is completely non-singular. It is
characterized by two independent positive parameters,   $(M,N)$.  A particular
solution in this class, $N = M^2$ is  the so-called    $a=1$  extremal magnetic
black hole  of Gibbons \cite{Gib}. It   is supersymmetric when embedded into
pure ${\cal N}=4$ supergravity without vector multiplets \cite{US}.
The absence of singularities in stringy frame for this solution
 was observed  previously in \cite{GHS} where the configuration was referred to
as a   ``bottomless hole".
Now we have a more general solution with analogous properties which
for $r \rightarrow 0$  approaches a throat
with a radius proportional to the norm of the magnetic
charges:
\begin{equation}
ds^2 \rightarrow - dt^2 +  d\eta^2  + 4N\, d\Omega^2\  ,
\end{equation}
where $d\eta^2 =4N (d \ln r)^2$. The
dilaton tells us that inside the throat the theory is in the strong coupling
regime. The scalars ${\cal M}$ are non-singular.
We have plotted the stringy  scalar curvature for some values of positive $M $
and $N  $, see Fig. 1.

In the electric counterpart of this solution we have a singularity
at $r=0$ in string metric but we are in the weak coupling regime. All solutions
in this class are known to have the non-extremal black hole partners with the
singularity covered by the horizon \cite{Sen}.

\item $M\geq 0, \, N= 0 $ {\it extremal supersymmetric  magnetic black holes,
singular
in stringy frame}.

We have plotted the values of the stringy scalar curvatures
for few values of the mass, see Fig. 2. The magnetic charge of the graviphotons
equals the magnetic charge of the
vector multiplets.
This one-parameter solution with $N=0,  M > 0 $ corresponds to the $a=\sqrt 3$
extremal limit of the magnetic black hole of Gibbons and Perry \cite{GP}. This
solution is singular
at $r=0$ in the four-dimensional space in canonical as well as in the stringy
frame. For the vanishing mass $M$ the solution becomes trivial.

\item $ M \geq 0, \, N < 0$ {\it supersymmetric massive and massless magnetic
white holes (repulsons)}.

When the magnetic charge of the vector multiplets exceeds the graviphoton
charge an additional  singularity appears at
$
r_0= \sqrt 2 (P_{\rm vec} - P_{\rm gr}) > 0
$.
We have discussed the nature of this singularity  in \cite{K} where the
corresponding singularities  in canonical frame were shown to have  reflecting
properties. For this reason such solutions cannot be called black holes. One
may call them either white holes or repulsons. The curvature
for the magnetic solutions in stringy frame with an additional singularity is
plotted in Fig. 3. Note that the dilaton (the string coupling) becomes very
small near the singularity at
$r=r_0$.
In the massless case the graviphoton charge vanishes, however, the charge of
the vector multiplets
is non-vanishing. We have a one-parameter massless monopole solution which is
singular at
$
r_0=\sqrt 2  P_{\rm vec} > 0
$.
Solutions with $N<0$   do not correspond to the extreme limit of any known
black holes.

\end{itemize}

One can easily derive the electric multi-center solutions from our
multi-monopole solutions in an O$(6,22)$-symmetric form. In the stringy frame
we
get
\begin{equation}   \label{electr}
ds^2_{\rm str}= -  e^{ 4 U}dt^2 +  d\vec{x}^2 \ , \qquad   \phi=U \ ,
 \qquad E_i ^{(a)}  = e^{4 U} ({\cal M} L)_{ab}\, \partial_i \chi^b\ ,
\end{equation}
where $U$ and ${\cal M}$ are defined in eq. (\ref{monopole}). The electric
fields
$E_i ^{(a)} = F_{ti}^{(a)}$ also form an O$(6, 22)$ vector and
the norm of the electric and magnetic
charges are equal, $|Q|=|P|$.
We will identify these electric configurations with the states in the string
spectrum
(for details see \cite{Sen}, \cite{DuffRum}). Since our solution saturates the
Bogomolny
bound (\ref{bogom}) the right-moving oscillator modes have
$N_R = \frac{1}{2}$. For the left-moving part we obtain
\begin{equation}
N_L - 1 = {1 \over 2} (Q_{\rm gr}^2 - Q_{\rm vec}^2) = {1\over 2}  (P_{\rm
gr}^2 - P_{\rm vec}^2) =  N \ .
\end{equation}
Thus,  the solution (\ref{electr}) describes the following states:

\noindent 1) $N_L=0$, $N_R=\frac{1}{2} $  massive  and massless white holes
($N=-1$);
 the massless configuration can be thought of  as a ground state of the theory.
 \newline
2) $N_L=1$, $N_R=\frac{1}{2}$ extremal $a=\sqrt{3}$ black holes ($N=0$).
\newline
3) $N_L \geq 2$, $N_R=\frac{1}{2}$ discrete set of extremal black holes
   ($N \geq 1$; for $N =  M^2 =$ $N_L -1$ they reduce to  $a=1$ black holes).

Our magnetic configurations (\ref{monopole}) appear now as solitons
related to  these elementary string states via S-duality.

It was suggested recently \cite{Tsey}  that one can smear the singularities at
$r=0$ of the spherically symmetric  backgrounds of string theory by introducing
the sources at the origin. This proposal requires some further study. It may
solve the problem with singularities at $r=0$ for $M>0, N\geq 0$ solutions in
all frames, since also the dilaton would become non-singular at
$r=0$. Still the solutions with $M\geq 0, N<0$ would have the singularity at
$r=r_0> 0$ when considered as the solutions of the four-dimensional geometry.
The uplifted geometry, however,  may be non-singular.

Stringy $\alpha'$-corrections are known to modify the uplifted solution, when
considered as embedded into the ten-dimensional geometry. To avoid Lorentz and
supersymmetry anomalies one has to supplement the configuration with  some
non-Abelian fields via embedding of the spin connections into the gauge group.
The symmetry of the solution will become O(6,6)$\times$G  where G is some
non-Abelian group, which is part of $E_8 \times E_8$ or SO(32). We hope to
study these issues in future.

\section*{Acknowledgements}
The work of K.B. is supported by the DFG and by a grant of
the DAAD. He would also like to thank the Physics Department
of the Stanford University for its hospitality.
The work of  R.K. is  supported by  the NSF grant PHY-8612280.

\ifx\nopictures Y\else  \input epsf  \fi

\begin{figure}[t]
\begin{minipage}[t]{8.3cm}
\ifx\nopictures Y\else \centerline{\hskip - .5 cm
\epsfxsize 3.4in \epsfbox{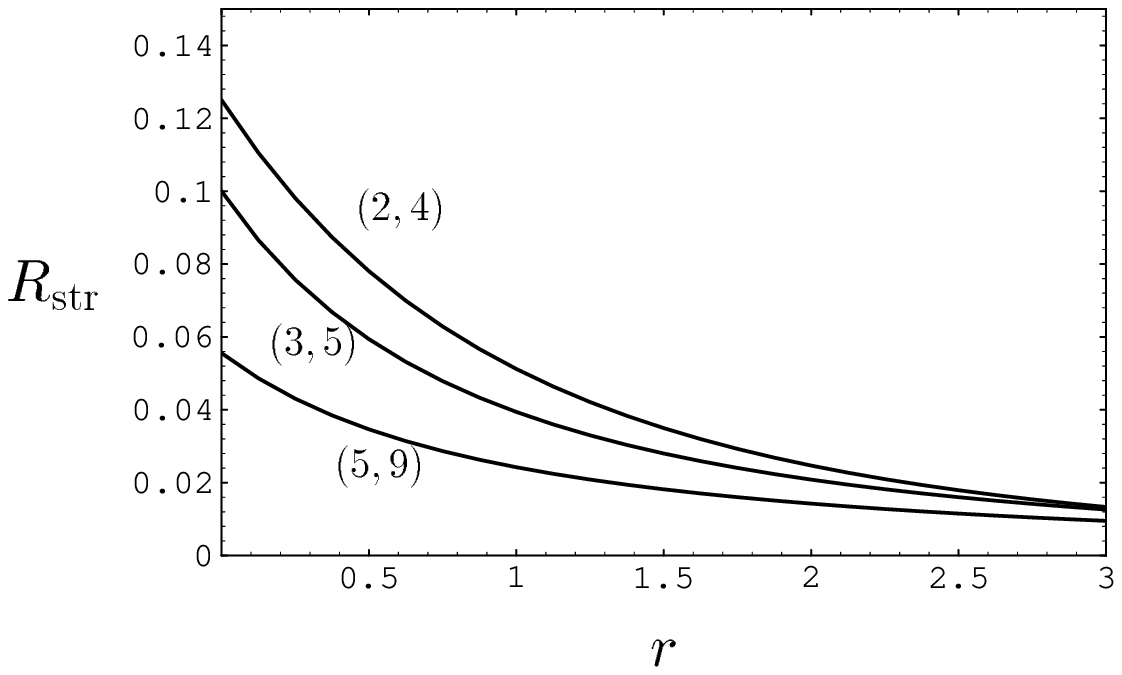}} \fi
\caption{The stringy  scalar curvature  of \mbox{$M>0$}, $N >0$
monopoles as the function of $r$. Each curve
is labeled with its $(M, N)$ pair. The upper curve corresponds to $a^2=1$
extremal
magnetic black hole with $N= M^2$. All solutions have finite curvature
everywhere.}

\label{F1}
\end{minipage}
\hfill
\begin{minipage}[t]{8.3cm}


\ifx\nopictures Y\else \centerline{\hskip - .3 cm
\epsfxsize 3.4in \epsfbox{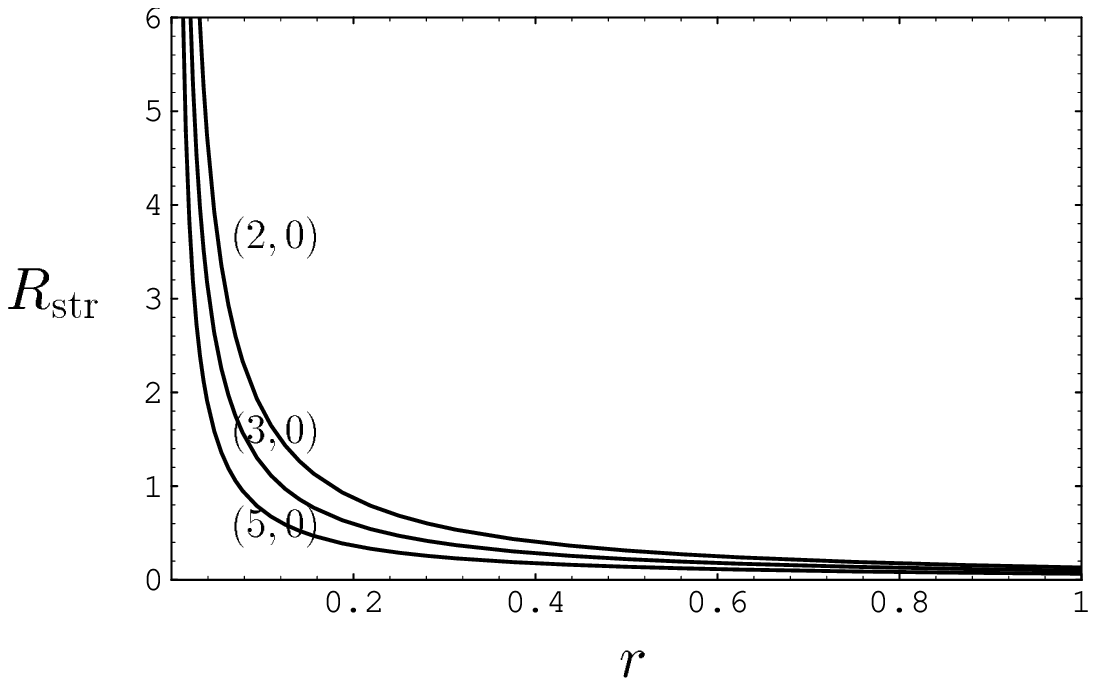}} \fi
\caption{The stringy  scalar curvature  of \mbox{$M>0$},
$N =0$ monopoles  ($a^2= 3$ extremal magnetic black holes). Each curve
is labeled with its $(M,0)$ pair. The curvature is infinite at $r=0$.}
\label{F2}
\end{minipage}
\end{figure}

\begin{figure}
\ifx\nopictures Y\else \centerline{\hskip - .8 cm
\epsfxsize 3.4in \epsfbox{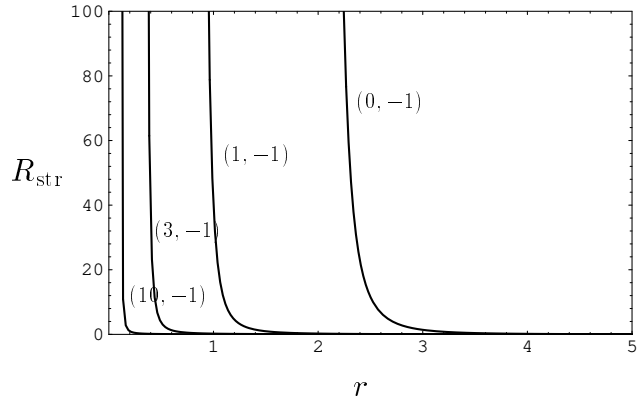}} \fi
 \vskip 0.2cm
\caption{The stringy  scalar curvature  of  $M\geq 0, \, N =-1 $ monopoles.
Each
curve
is labeled with its $(M,N)$ pair. The curvature is infinite at $r_0 =2
[(M^2+1)^{1/2}-M]$. The  massless  magnetic configuration (the curve (0,-1))
 is singular at $r_0 =2 $.}
\label{F3}
\end{figure}

\end{document}